\documentclass[runningheads,aps,tighten,epsf,epsfig,rotate,cite]{svmult}
\usepackage{makeidx}   
\usepackage{graphicx}  
\usepackage{subeqnar}  
\usepackage{multicol}  
\usepackage{cropmark} 
\usepackage{physprbb}  
\newcommand{\be}{\begin{equation}}
\newcommand{\ee}{\end{equation}}
\newcommand{\bea}{\begin{eqnarray}}
\newcommand{\eea}{\end{eqnarray}}

\def\gappeq{\mathrel{\rlap {\raise.5ex\hbox{$>$}}
{\lower.5ex\hbox{$\sim$}}}}

\begin{document}
\title*{Is Nothing Sacred?\\
Vacuum Energy, Supersymmetry and Lorentz Breaking from
Recoiling $D$ branes}
\toctitle{Is Nothing Sacred?
\protect\newline Vacuum Energy, Supersymmetry and Lorentz Breaking from
Recoiling $D$ branes}
%
%
\titlerunning{Is Nothing Sacred?\\
Vacuum Energy, Supersymmetry and Lorentz Breaking from
Recoiling $D$ branes}
%
\author{John Ellis\inst{1}
\and N.E. Mavromatos\inst{2}
\and D.V. Nanopoulos\inst{3,}\inst{4,}\inst{5}}

\authorrunning{John Ellis et al.}
%
%
\institute{Theory Division, CERN, CH-1211 Geneva 23, Switzerland
\and Theoretical Physics Group, Department of Physics,
     King's College London, Strand, London WC2R 2LS, UK
\and Department of Physics, Texas A \& M University, 
College Station, TX~77843-4242, USA 
\and Astroparticle Physics Group, Houston
Advanced Research Center (HARC), 
Mitchell Campus,
Woodlands, TX 77381, USA
\and Academy of Athens, Chair of Theoretical Physics, 
Division of Natural Sciences, 
28~Panepistimiou Avenue, 
Athens 10679, Greece}

\titlerunning{Is Nothing Sacred? ...}
\maketitle              

\begin{center}
ACT-6-00, \,\,\,\, CERN--2000--136, \,\,\,\, CTP-TAMU--12/00,
\,\,\,\, gr-qc/0005100
\end{center}

\begin{abstract}
Classical superstring vacua have zero vacuum energy and  are
supersymmetric and Lorentz-invariant. We argue that all
these properties may be destroyed when quantum aspects of the
interactions between particles and non-perturbative vacuum
fluctuations are considered. A toy calculation of string/$D$-brane
interactions using a world-sheet approach indicates that quantum
recoil effects - reflecting the gravitational back-reaction on
space-time foam due to the propagation of energetic particles -
induce non-zero vacuum energy that is linked to supersymmetry breaking
and breaks Lorentz invariance. This model of space-time foam also
suggests the appearance of microscopic event horizons.

\end{abstract}

\section{Introduction}

String theory is our best (only?) hope for a consistent quantum theory of
gravity and all the other particle interactions~\cite{SR}. 
As such, it should
address the non-perturbative nature of the quantum-gravitational
vacuum~\cite{polch}. 
On basic physical grounds, one would expect the vacuum to exhibit
non-perturbative quantum fluctuations, much as the QCD vacuum exhibits
non-perturbative topological fluctuations such as instantons. In the case
of quantum gravity, the Planck-scale vacuum fluctuations go by the
picturesque name of space-time foam~\cite{foam}, 
which string theory should have the
ambition to understand.  However, many extant string approaches are
limited to the identification of classical string vacua~\cite{SR,polch}, 
and the
consideration of the interactions of particles propagating through them. 

Certain approaches to quantum gravity, such as the loop 
description~\cite{ashtekar}, do
attempt to understand certain quantum aspects of space-time foam.  We have
been making~\cite{emn98} parallel attempts in an approach to string theory that
introduces on the world sheet a Liouville field~\cite{david88,distler89}, 
which is identified with
a renormalization scale and the time coordinate $t$.  This approach
accommodates departures from conventional critical string theory, as may
be needed to describe transitions between different classical vacua and
other non-perturbative phenomena. In this connection, it is encouraging to
learn~\cite{EMND} that $D$ branes appear naturally within this world-sheet
Liouville
approach to non-critical string theory.

Several of the sacred properties of classical string theory may be
unsustainable in such an approach. We have argued that it provides a
natural description for a system of particles interacting with $D$ branes,
which we use to
model non-perturbative quantum fluctuations in the space-time background.
As a consequence of the induced non-criticality, the vacuum energy becomes
non-zero~\footnote{We recall that non-perturbative QCD effects
also change the vacuum energy.}.  It is, however, not constant: in our
approach~\cite{ellis98,emncosmol}, 
there is a component in the vacuum energy which relaxes towards
zero $\propto 1 / t^2$. If this component is dominant today, the vacuum
energy is not a `cosmological constant'. 

In the context of a supersymmetric theory, non-zero vacuum energy carries
with it the stigma of supersymmetry breaking. Indeed, as we demonstrate
explicitly within our approach~\cite{emncosmol,adrian+mavro99,am}, 
quantum-gravitational recoil does
break supersymmetry. Whether this can be related to the supersymmetry
breaking required by the non-observation of supersymmetric partners of the
known particles is, however, a weighty subject beyond the scope of this
talk.

Non-critical string theory also carries with it the stigma of a breakdown
of Lorentz invariance. We are not (sufficiently?) shocked by this
corollary. Indeed, our $D$-brane model is not the only treatment of
space-time foam in which the vacuum may acquire non-trivial optical
properties
such as a refractive index, birefringence and diffusive
particle propagation~\cite{aemn,sarkar,pullin}. 
Any and all of these properties violate
Lorentz invariance~\cite{foam,mestres,lv}.

The structure of this talk is as follows. In Section 2 we review our
description of $D$-brane interactions and recoil, arguing that they lead
in general
to departures from criticality that can be accommodated within a
world-sheet Liouville approach. 
We discuss in Section 3 recoil-induced deformations of the
background space-time metric. 
We show in Section 4 how $D$-brane recoil
may induce a non-zero but time-dependent contribution to the vacuum
energy, and also a cosmic expansion of our world,
viewed as a brane embedded in a higher-dimensional (bulk) space-time. 
The appearance of a microscopic horizon in this approach to space-time
foam is described in section 5, and
the breaking of supersymmetry by such a recoil effect is discussed
in Section 6. We discuss in Section 7 some observable consequences
of our recoil formalism, namely a
non-trivial
refractive index {\it in vacuo}, as well as diffusive particle
propagation. We provide in Section 8 a brief summary of observational
constraints on such possible space-time foam signatures of quantum
gravity,
and Section 9 discusses the outlook for our approach. 

\section{A Review of the World-Sheet Description of $D$-brane
Interactions}

We first review the world-sheet formalism for treating
$D$-brane interactions~\cite{kogan96,mavro+szabo,ellis96,ellis98}.
In this approach, the recoil of a
$D$-brane when struck by a closed-string state or by another $D$-brane
is described mathematically by a logarithmic conformal field
theory~\cite{lcft}. Such theories
lie on the border between finite conformal field theories and
general renormalizable two--dimensional quantum field theories. As such,
they provide the first stepping-stone away from the conformal field
theories used to characterize critical string theories. They are
relevant~\cite{kogan96,mavro+szabo,ellis96} to this
problem because the recoil process involves a change of state
(transition) in the string background, which cannot be
described by a conformal field theory. In the language of the
world-sheet, this change of state
induced by the recoil process can be described as a change in the
$\sigma$-model background, and as such is a non-equilibrium process.
This property is reflected~\cite{ellis96,mavro+szabo} in the logarithmic
operator algebra itself.

\begin{figure}[b]
\begin{center}
\includegraphics[width=.3\textwidth]{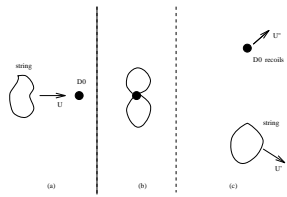}
\end{center}
\caption[]{Schematic representation of the scattering
of a closed-string state on a  $D$ particle ($D0$ brane)
embedded in the target space-time.
(a) The closed-string state,
moving with velocity $U$ along a given spatial direction, (b)
strikes at a given instant in time the $D0$ brane,
which then (c)
recoils, distorting the space-time around it.}
\label{fig1}
\end{figure}

As discussed in references~\cite{kogan96,ellis96,mavro+szabo}, the
recoil of a $D$-brane string soliton after
interaction with a closed-string state, illustrated in Fig.~\ref{fig1}, is
characterized
by a $\sigma$ model deformed by a pair of logarithmic
operators~\cite{lcft}:
\begin{equation}
C^I_\epsilon = \epsilon \Theta_\epsilon (X^I),\qquad
D^I_\epsilon = X^I \Theta_\epsilon (X^I), \qquad I \in \{0,\dots, 3\}
\label{logpair}
\end{equation}
defined on the boundary $\partial \Sigma$ of the string world
sheet. Here $X^I, I\in \{0, \dots, p\}$ obey Neumann boundary
conditions on the string world sheet, and denote the $D$-brane
coordinates, whilst $\epsilon\rightarrow0^+$ is
a regulating parameter and $\Theta_{\epsilon}(X^{I})$ is a
regularized Heaviside step function.
The remaining $y^i, i\in \{p+1, \dots, 9\}$ in (\ref{logpair}) denote the
transverse bulk directions.

In the case of $D$ particles~\cite{kogan96,ellis96,mavro+szabo},  
the index $I$ takes the
value $0$ only, in which case the operators (\ref{logpair}) act as
deformations of the conformal field theory on the world sheet. The
operator
\begin{equation}
u_i \int _{\partial \Sigma} \partial_n X^i D_\epsilon
\label{movement}
\end{equation}
describes the movement of the $D$ brane induced by the scattering,
where $u_i$ is its recoil velocity, and $y_i \int _{\partial
\Sigma} \partial_n X^i C_\epsilon $ describes quantum fluctuations
in the initial position $y_i$ of the $D$ particle. It has been
shown rigorously~\cite{mavro+szabo} that the logarithmic conformal
algebra ensures energy--momentum  conservation during the recoil
process:
\begin{equation}
u_i = M_D ( k^1_i + k^2_i), 
\label{conservation}
\end{equation}
where $k^1 (k^2)$ is
the momentum of the propagating closed string state before (after)
the recoil, and $M_D=1/(\ell _s g_s)$ is the mass of the $D$ brane,
where $g_s$ is the string coupling, which is assumed
here to be weak enough to ensure that the $D$ brane is very massive,
and $\ell _s$ is the string
length.

The second member of the logarithmic pair of $\sigma$-model 
deformations is
\begin{equation}
y_{i} \int _{\partial \Sigma} \partial_n X^i C_\epsilon~,
\label{logpair2}
\end{equation}
where, in order to realize the logarithmic
algebra between the operators $C$ and $D$, one uses
as a regulating parameter~\cite{kogan96}
\begin{equation}
\epsilon^{-2} \sim \ln [L/a] \equiv \Lambda,
\label{defeps}
\end{equation}
where $L$ ($a$) is an infrared (ultraviolet) world--sheet cutoff.
The recoil operators (\ref{logpair2}) are
relevant, in the sense of the renormalization group for the
world--sheet field theory, having small conformal dimensions
$\Delta _\epsilon = -\epsilon^2/2$. Thus the $\sigma$-model
perturbed by these operators is not conformal for $\epsilon \ne
0$, and the theory requires Liouville
dressing~\cite{david88,distler89,ellis96}. The consistency of this
approach is supported by the proof of momentum conservation
during the scattering process~\cite{mavro+szabo}.

The world-sheet renormalization-group $\beta$ functions of the 
relevant recoil couplings $g_{Ii}^D \equiv g_{Ii},~I\in\{0, \dots,
p\}, ~i\in\{p+1, \dots, 9\}$ have the form
\begin{equation}
  \beta_{g_{Ii}} = \frac{d}{d t} g_{Ii} = -\frac{1}{2t} g_{Ii} , \qquad
t \sim \epsilon ^{-2}~.
\label{betaf}
\end{equation}
Thus, one may construct an exactly marginal set of
couplings ${\overline g_{Ii}}$ by redefining
\begin{equation}
{\overline g_{Ii}} \equiv \frac{g_{Ii}}{\epsilon}~.
\label{marginal}
\end{equation}
The renormalized couplings ${\overline g_{0i}}$ were 
shown~\cite{mavro+szabo} to play the r\^ole of the physical recoil
velocity of the $D$ brane, whilst the remaining ${\overline
g_{Ii}},~I\ne 0$, describe the folding of the $Dp$ brane for
$p\ne 0$~\cite{emw99}. Here we assume, generalizing~\cite{mavro+szabo}, 
that the (bare) recoil couplings for all
$I$ are of equal strength and related to the transverse momentum
transfer as
\begin{equation}
  g_{Ii} =g_s \frac{\Delta P_i}{M_s}~, I=0, \dots ,m, ~ i=m+1, \dots D
\label{momtransf}
\end{equation}
for a $D$ brane embedded in a $D$-dimensional space-time.

An important technical remark~\cite{kogan96} is now in order:
for reasons of convergence of the world-sheet
path integral, the Neumann coordinate $X^0$ must be
Euclideanized. It is only in this case that
the identification (\ref{defeps}), with $\epsilon^2 >0$,
leads to a mathematically
consistent logarithmic algebra of operators.
This can be understood simply by the fact that,
in the pertinent world-sheet computations
of correlation functions of logarithmic operators
(\ref{logpair}),
one encounters~\cite{kogan96}
the free propagator of the Neumann coordinates $X^I$:
\be
{\cal G}_0 = {\rm \lim}_{\sigma \rightarrow 0}<X^I(\sigma) X^J
(0)>_* \sim \eta_{IJ} \ln [L/a] \label{freeprop}
\ee
where $< \cdots >_*$ denotes a world-sheet expectation value
calculated using the
free-string world-sheet action on a flat target
space-time manifold $\{ X^I \}$,  and $\eta^{IJ}$ is the target-space
metric. For Euclidean world sheets, one takes
$\eta^{IJ}=\delta^{IJ}$, which is essential for the convergence
of world-sheet path-integral expressions entering in the
respective correlators. 

In our picture, we view the $(3+1)$-dimensional 
physical world as a brane.
This Euclideanization implies that
the (longitudinal) Neumann coordinates define a $D4$ domain
wall in the bulk space-time, which, after analytic continuation of
the coordinate $X^0$, will result in our four-dimensional
space-time.
However, the analytic continuation takes place
only at the very end of the calculations. This is
important for our subsequent discussion, and should always be
understood in what follows.

\section{Deformations of Space-Time Induced by $D$-Brane Recoil}

As discussed in~\cite{ellis96,kanti98}, the deformations
(\ref{logpair}) create a local distortion of the space-time
surrounding the recoiling folded $D$-brane, which may be
determined using the method of Liouville dressing.
In~\cite{ellis96,kanti98} we concentrated on describing the
resulting space-time in the case when a $D$-particle defect embedded in
a $D$-dimensional space-time recoils after the scattering of a
closed string. To leading order in
the recoil velocity $u_i$ of the $D$ particle, the resulting
space-time was found, for times $t \gg 0$ long after the
scattering event at $t=0$, to be equivalent to a Rindler wedge,
with apparent `acceleration' $\epsilon u_i$~\cite{kanti98}, where
$\epsilon$ is defined above (\ref{defeps}).
For times $t < 0$, the space-time is flat Minkowski~\footnote{There is
hence a discontinuity at $t =0$, which leads to particle
production and decoherence for a low-energy spectator field theory
observer who performs local scattering experiments
long after the scattering, and far away from the
location of the collision of the closed string with the
$D$ particle~\cite{kanti98}.}.

\begin{figure}[b]
\begin{center}
\includegraphics[width=.3\textwidth]{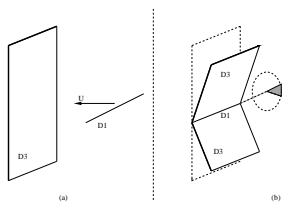}
\end{center}
\caption[]{Schematic representation of the folding effect
in $D$-brane/$D$-brane collisions:
(a) a $D1$ brane
moving with velocity $U$ along a `bulk' direction
perpendicular to a
$Dp$ brane embedded in a
$D$-dimensional 
space time strikes the $Dp$ brane (b), which is then
folded, and the space-time around it is distorted. 
The dashed circle around the $D1$ direction in
(b) indicates the angular deficit that appears when the bulk
direction along which the $D1$ brane was moving is compactified to
a circle. A generalization to a higher-dimensional case for the
incident brane is straightforward. In that case the deficit (in
the compact case) is a higher-dimensional solid hyperangle.}
\label{fig2}
\end{figure}


This situation is easily generalized to $Dp$ branes~\cite{emw99}
as seen in Fig.~\ref{fig2}.
The folding/recoil deformations of the $Dp$ brane (\ref{logpair2})
are relevant deformations, with anomalous dimension
$-\epsilon^2/2 $, which disturbs the conformal invariance of the
world-sheet $\sigma$ model, and restoration of conformal invariance
requires
Liouville dressing~\cite{david88,distler89,ellis96}, as discussed above.
To determine the effect of
such dressing on the space-time geometry, it is essential  to
write~\cite{ellis96} the boundary recoil deformations as  bulk
world-sheet deformations
\begin{equation}
\int _{\partial \Sigma} {\overline g}_{Iz} x\Theta_\epsilon (x)
\partial_n z =
\int _\Sigma \partial_\alpha \left({\overline g}_{Iz} x\Theta_\epsilon (x)
\partial ^\alpha z \right)
\label{a1}
\end{equation}
where the ${\overline g}_{Iz}$ denote the renormalized
folding/recoil couplings (\ref{marginal}), in the sense discussed
in~\cite{mavro+szabo}. As we have already mentioned, such
couplings are marginal on a flat world sheet.
The operators (\ref{a1}) are marginal also on a curved
world sheet, provided~\cite{distler89} one dresses the (bulk)
integrand by multiplying it by a factor $e^{\alpha_{Ii}\phi}$,
where $\phi$ is the Liouville field and $\alpha_{Ii}$ is the
gravitational conformal dimension, which is related to the
flat-world-sheet anomalous dimension $-\epsilon^2/2$ of the recoil
operator, viewed as a bulk world-sheet deformation, as
follows~\cite{distler89}:
\begin{equation}
\alpha_{Ii}=-\frac{Q_b}{2} +
\sqrt{\frac {Q_b^2}{4} + \frac {\epsilon^2}{2} }
\label{anom}
\end{equation}
where $Q_b$ is the central-charge deficit of the bulk world-sheet
theory. In the recoil problem at hand, as discussed
in~\cite{kanti98},
\be
Q_b^2 \sim \epsilon^4/g_s^2  > 0
\label{centralcharge}
\ee
for weak folding deformations $g_{Ii}$, and hence one is
confronted with a {\it supercritical} Liouville theory. This
implies a {\it Minkowskian-signature} Liouville-field kinetic term
in the respective $\sigma$ model~\cite{aben89}, which prompts one
to interpret the Liouville field as a time-like target-space
field. In our context, this will be a {\it second} time
coordinate~\cite{emn98}, which is independent of the
(Euclideanized) $X^0$. The presence of this second `time'
does not affect physical observables, which are defined
for appropriate slices with fixed Liouville coordinate, e.g., $\phi
\rightarrow \infty$, or equivalently
$\epsilon \rightarrow 0$.
From the expression (\ref{centralcharge}) we conclude (cf.
(\ref{anom})) that $\alpha_{Ii} \sim \epsilon $ to leading order
in perturbation theory in $\epsilon$, to which we restrict
ourselves here.

We next remark that, as the analysis of~\cite{ellis96} indicates,
the $X^I$-dependent field operators
$\Theta_\epsilon (X^I)$ scale as follows with $\epsilon$:
$\Theta_\epsilon(X^I) \sim e^{-\epsilon X^I}
\Theta(X^I)$, where $\Theta(X^I)$ is a Heavyside step function
without any field content, evaluated in the limit $\epsilon \rightarrow 0^+$.
The bulk deformations, therefore, yield the following
$\sigma$-model terms:
\begin{equation}
\frac{1}{4\pi \ell_s^2}~\int _\Sigma 
\sum_{I=0}^{3} \left( \epsilon^2 {\overline g}^C_{Ii} + \epsilon 
{\overline g}_{Ii} X^I\right)
e^{\epsilon(\phi_{(0)} - X^I_{(0)})}\Theta(X^I_{(0)})
\partial_\alpha \phi 
\partial^\alpha y_i~
\label{bulksigma}
\end{equation}
where the subscripts $(0)$ denote world-sheet zero modes, and 
${\overline g}^C_{0i}=y_i$.

Upon the interpretation of the Liouville zero mode $\phi_{(0)}$ as
a (second)
time-like coordinate, the deformations (\ref{bulksigma}) yield
metric
deformations of the generalized space-time
with two times. The metric components
for fixed Liouville-time slices can be
interpreted in~\cite{ellis96}
as expressing the distortion of the space-time
surrounding the recoiling $D$-brane soliton.

For clarity,
we now drop the subscripts $(0)$ for the rest of this paper,
and we work in a region of space-time
such that $\epsilon (\phi - X^I)$ is finite
in the limit $\epsilon \rightarrow 0^+$.
The resulting space-time distortion is therefore
described by the metric elements
\begin{eqnarray}
&~& G_{\phi\phi} = -1, \qquad G_{ij} =\delta_{ij}, \qquad
G_{IJ}=\delta_{IJ}, \qquad G_{iI}=0,   \nonumber \\
&~& G_{\phi i} = \left(\epsilon^2 {\overline g}^C_{Ii} +
 \epsilon {\overline g}_{Ii}X^I \right)\Theta (X^I)~,
\qquad i=4, \dots 9,~~I=0, \dots 3
\label{gemetric}
\end{eqnarray}
where the index $\phi $ denotes Liouville `time', not to be confused
with the Euclideanized time which is one of the $X^I$.
To leading order in $\epsilon {\overline g}_{Ii}$,
we may ignore the $\epsilon^2 {\overline g}^C_{Ii}$ term.
The presence of $\Theta(X^I)$ functions and
the fact that we are working in the region $y_i >0$
indicate that
the induced space-time is piecewise continuous~\footnote{The
important implications for non-thermal particle production
and decoherence for a spectator low-energy field theory
in such space-times were discussed in~\cite{kanti98,ellis96}, where
the $D$-particle recoil case was considered.}.
In the general recoil/folding case considered in this article,
the form of the resulting patch of the surrounding
space-time can be determined fully if one computes
the associated curvature tensors, along the lines
of~\cite{kanti98}.

We next study in more detail some physical aspects of
the metric (\ref{gemetric}),
restricting ourselves, for simplicity, to the case
of a single Dirichlet dimension $z$ that
plays the r\^ole of a bulk dimension
in a set up where there are 
Neumann coordinates $X^I$, $I=0,\dots3$
parametrizing a D4 (Euclidean) brane, interpreted as
our four-dimensional space-time.
Upon performing the time transformation
$\phi \rightarrow \phi - \frac{1}{2}\epsilon {\overline g}_{Iz} X^I z $, the
line element (\ref{gemetric}) becomes:
\begin{eqnarray}
&~&ds^2 =-d\phi^2 + \left(\delta_{IJ}
-\frac{1}{4}\epsilon^2{\overline g}_{Iz}{\overline g}_{Jz}~z^2\right)~dX^I dX^J
+  
\nonumber \\
&~& \left(1 + \frac{1}{4}\epsilon ^2
{\overline g}_{Iz}{\overline g}_{Jz}~X^I~X^J\right)~dz^2 -
\epsilon {\overline g}_{Iz}~z~dX^I~d\phi~, \nonumber \\
\label{bendinglineel}
\end{eqnarray}
where $\phi$ is the Liouville field which, we remind the reader,
has Minkowskian signature in the case of supercritical
strings that we are dealing with here.

One may now make a general coordinate transformation on the
brane $X^I$ that diagonalizes the pertinent induced-metric
elements in (\ref{bendinglineel})~\footnote{Note that general
coordinate invariance is assumed to be a good symmetry on the
brane, away from the `boundary' $X^I=0$.}. For instance, to
leading order in the deformation couplings ${\overline
g}_{Iz}{\overline g}_{Jz}$, one may redefine the $X^I$ coordinates
by 
\begin{eqnarray} X^I &\rightarrow& X^I -\frac{\epsilon^2}{8}z^2
{\overline g}_{Iz} \sum_{J \ne I}{\overline g}_{Jz}X^J,\nonumber\\
z &\rightarrow& z \left(1 + \frac{\epsilon^2}{8}\sum_{I \ne J}
{\overline g}_{Iz}{\overline g}_{Jz}X^{I}X^{J}\right)
\end{eqnarray}
which leaves only diagonal elements  of the
metric tensor on the (redefined) hyperplane $X^I$. In this case,
the metric becomes, to leading order in $g_{Iz}^2$ and
in the case where $
\epsilon {\overline g}_{Iz}z << 1$:
\begin{eqnarray}
&~&ds^2 =-d\phi^2 + \left(1
-\alpha^2 ~z^2\right)~(dX^I)^2
+ \left(1 + \alpha^2 ~(X^I)^2\right)~dz^2 - \epsilon {\overline g}_{Iz}~z~dX^I~d\phi~,
\nonumber \\
&~& \alpha=\frac{1}{2}\epsilon {\overline g}_{Iz} \sim g_s |\Delta P_z|/M_s
\label{bendinglineel3}
\end{eqnarray}
where the last expression makes it clear that, upon utilizing
(\ref{marginal}, \ref{momtransf}), one can express the parameter
$\alpha$ (in the limit $\epsilon \rightarrow 0^+$)
in terms of the (recoil) momentum transfer along the bulk direction.
As we see later on, this parameter is responsible for the mass hierarchy
in the problem, assuming that the string scale $M_s$ is close to
Planck mass scale $10^{18}$ GeV, for ordinary string-theory
couplings of order $g_s^2/2\pi =1/20$.

A last comment, which is important for our purposes here, 
concerns the case in which
the metric (\ref{bendinglineel3}) is {\it exact}, i.e., it holds
to all orders in ${\overline g}_{Iz}z$.
This is the case where
there is no world-sheet
tree-level momentum transfer. This naively corresponds to the case
of static intersecting branes. However, the whole philosophy of
recoil~\cite{kogan96,mavro+szabo} implies that, even in that case,
there are quantum fluctuations induced by the sum over genera of the
world sheet. The latter implies the existence of a statistical
distribution of logarithmic deformation couplings of Gaussian type
about a mean field value ${\overline g}^{c}_{Iz}=0$. Physically,
the couplings
${\overline g}_{Iz}$ represent recoil velocities of the intersecting
branes,
hence these Gaussian fluctuations
represent the effects of quantum fluctuations about the
zero recoil-velocity case, which may be considered as quantum
corrections to the static intersecting-brane case.
We therefore consider a statistical average
$<< \cdots >>$ of the line element (\ref{bendinglineel})
\begin{eqnarray}
&~&<<ds^2>> =-d\phi^2 + \left(1
-\frac{1}{4}\epsilon^2
<<{\overline g}_{Iz}{\overline g}_{Jz}>>~z^2\right)~dX^I dX^J
+ \nonumber \\
&~& \left(1 + \frac{1}{4}\epsilon ^2
 <<{\overline g}_{Iz}{\overline g}_{Jz}>>~X^I~X^J\right)~dz^2
- \epsilon <<{\overline g}_{Iz}>>~z~dX^I~d\phi~, \nonumber \\
\label{bendinglineel2}
\end{eqnarray}
where
\be
<< \cdots >>=\int _{-\infty}^{+\infty}d{\overline g}_{Iz}
\left(\sqrt{\pi}\Gamma \right)^{-1} 
 e^{-{\overline g}_{Iz}^2/\Gamma^2} (\cdots) \label{gauss}
\ee
and the width $\Gamma$ has been calculated
in \cite{mavro+szabo}, 
is found after summation over world-sheet
genera to be
proportional to the string coupling $g_s$. 
In fact~\cite{mavro+szabo}, it can be shown that $\Gamma$ scales 
as $\epsilon {\overline \Gamma}$, where ${\overline \Gamma}$ 
is independent of $\epsilon$. This will be important later on, when 
we consider the identification of $\epsilon$ with the target time $t$.

{\it This is the model of string-inspired space-time foam
used in this work. The quantum recoil
fluctuations reflect the response of the 
$D$ brane to the emission of virtual closed strings or other
branes. }

We see from (\ref{gauss}),
assuming that $g_{Iz}=|U_i|$ where $U_i=g_s \Delta P_i/M_s$
is the recoil velocity~\cite{kogan96,mavro+szabo}, that
the average line element
$ds^2$ becomes:
\begin{eqnarray}
&~&<<ds^2>> =-d\phi^2 + \left(1
-\alpha^2 ~z^2\right)~(dX^I)^2
+ \left(1 + \alpha^2 ~(X^I)^2\right)~dz^2,
\nonumber \\
&~& \alpha=\frac{1}{2\sqrt{2}}\epsilon^2 {\overline \Gamma}
\label{bendinglineel3ab}
\end{eqnarray}
The definition of $\alpha$ comes from evaluating the quantity
$<<{\overline g}_{Iz}^2>>$ using the statistical distribution (\ref{gauss}).
Thus the average over quantum fluctuations leads to a
metric of the form (\ref{bendinglineel3}), but with a parameter
$\alpha$ much smaller, being determined by the width (uncertainty)
of the pertinent quantum fluctuations~\cite{mavro+szabo}.
The metric (\ref{bendinglineel3ab})
is exact, in contrast to the metric (\ref{bendinglineel3})
which was derived for $z << 1/\alpha$. However, for our purposes
below we shall treat both
metrics as exact solutions of some string theory
associated with the recoil.

An important feature of the line element (\ref{bendinglineel3ab}) is
the existence of a {\it horizon} at $z=1/\alpha$ for {\it Euclidean}
Neumann coordinates $X^I$. Since the Liouville field
$\phi$ has decoupled after the averaging procedure, 
one may consider slices of this field, defined by $\phi$ = const, on
which  the physics of the observable world can be studied. From a
world-sheet renormalization-group view point this slicing
procedure corresponds to selecting a specific point in the
non-critical-string theory space. Usually, the infrared fixed
point
$\phi \rightarrow \infty$ is selected. In that case, from (\ref{defeps}),
one considers a slice for which $\epsilon^2 \rightarrow 0$.
But any other choice could do, so $\alpha$ may be considered
a small but arbitrary parameter of our effective theory.
The presence of a horizon raises the issue of how one
could analytically continue so as to pass to the space beyond the horizon.
The simplest way, compatible, as we shall show later with the low-energy
Einstein's equations, is to take the absolute value of $1-\alpha^2 z^2$
in the metric element (\ref{bendinglineel3}).

We next note that  
in~\cite{leonta}, where the above metric was initially derived as an
extension of the $D$--particle case of \cite{kanti98},
the metric was defined in all space $z \in R$
on a {\it slice} of the Liouville time $\phi$ = const by:
\begin{equation}
ds^2_{f} = \left|1
-\alpha^2 ~z^2\right|~(dX^I)^2
+ \left(1 + \alpha^2 ~(X^I)^2\right)~dz^2,
\label{bendinglineel3b}
\end{equation}
For small $\alpha$, which is the case studied here,
and for Euclidean
Neumann coordinates $X^I$, the scale factor in front of the
$dz^2$ term does not introduce any singular behaviour, and hence
for all qualitative purposes one may study the metric element~\cite{leonta}:
\begin{equation}
ds^2_{f} = \left|1
-\alpha^2 ~z^2\right|~(dX^I)^2
+ ~dz^2,
\label{bendinglineel4}
\end{equation}
which shares all the qualitative
features of the full metric (\ref{bendinglineel3b}), 
induced by the recoil process in the case of an uncompactified
`bulk' Dirichlet dimension $z$, as we consider
here~\footnote{For the case of compact dimension $z$
the situation changes drastically,
since in that case there are  
angular deficits
in the cycle around $z$~\cite{emw99}.}.  

Such a metric has been shown to satisfy Einstein's 
equations in the bulk, provided there is a non-trivial 
vacuum energy on the brane world at $z=0$, associated 
with the excitation due to the recoil~\cite{leonta}. 
An important point to note is that,
formally, our analysis leading to (\ref{bendinglineel3ab}) is valid
in the region of bulk space-time for which $z > 0$. However, one
may consider a {\it mirror} extension of the space-time to the
region $z < 0$, which we assume in this article. {}From now on,
therefore, we treat the metric (\ref{bendinglineel4}) as being
defined over the entire real axis for the bulk coordinate $z \in
R$. However, to make contact with the original recoil picture we
restrict ourselves to regions of space-time for which
$X^I >0$.

Non-trivial physics is also obtained if, following the 
spirit of \cite{emn98}, one {\it identifies}, in the expression 
for the metric (\ref{bendinglineel3ab}), the Liouville 
mode $\phi$ with the target time $X^0$. In that case, 
despite the fact that the coordinates $X^I$ are {\it Euclidean},
one obtains a Minkowskian signature for the (Liouville) time. 
Notice that the Liouville time now is not fixed, in contrast to the 
case considered in \cite{leonta}, but `runs' along 
generalized world-sheet renormalization-group trajectories 
of the associated Liouville string. 

In this case the metric element (\ref{bendinglineel3ab}) 
does not satisfy the classical Einstein's equations in the bulk,
but the `off-shell' Liouville conditions
for the $\sigma$-model couplings $\{ g^i \}$, corresponding to 
$\sigma$-model backgrounds of 
graviton, dilaton and appropriate matter fields in the low-energy 
limit of the associated string theory~\cite{distler89,emn98}:
\be
{\ddot g}^i + Q {\dot g}^i = -{\tilde \beta}^i 
\label{liouville} 
\ee
where the dot denotes a derivative with respect to the 
(world-sheet zero mode of the) Liouville field, $d/d\phi$, 
and the ${\tilde \beta}^i $ are the appropriate Weyl anomaly coefficients
on a flat world-sheet~\cite{emn98}, which are related to the ordinary 
world-sheet renormalization $\beta$ functions by
${\tilde \beta}^i = \beta^i + \delta g^i$. The quantities 
$\delta g^i$ denote terms expressing appropriate target-space
diffeomorphism variations of the fields $g^i$. The presence
of ${\tilde \beta}^i$ instead of $\beta^i$ is necessitated by the 
target-space diffeomorphism invariance of the 
$\sigma$ model~\footnote{Despite the fact that, eventually, 
it will be broken
by the recoiling background, this 
symmetry may be assumed formally valid 
for the $\sigma$ models at hand.}.  
 
In the case of \cite{leonta}, where a slice at an infrared fixed point 
$\phi \rightarrow \infty$ of the Liouville mode 
was considered, both sides of (\ref{liouville}) {\it vanish},
corresponding to the independence of the couplings/fields  $\{ g^i \}$
{}from the Liouville mode, and the satisfaction of Einstein's and other 
field equations derived from a low-energy string effective action.

In the present case, the background (\ref{bendinglineel3ab}) 
is constrained to satisfy (\ref{liouville}), with the field 
$\phi$ identified as the target-space time $X^0$ in the solution 
of (\ref{liouville}).  This needs a careful consistency check.
In general, the problem of solving (\ref{liouville}) in 
an arbitrary number of target space dimensions is very complicated.
Even in toy models of Liouville 
cosmology with two target space-time dimensions~\cite{diamandis} the
solutions of such equations, with
the Liouville mode being identified with the target time $X^0$, 
are but partially known, and only numerically at present.

However, the situation may become simpler
in certain regions of the bulk space time. Indeed,
far away from the horizon at $|z|=1/\alpha$, 
i.e., for $\alpha ^2 z^2 << 1$, 
the line element corresponding to the space-time 
(\ref{bendinglineel3ab}) after the identification 
$\phi =X^0$ becomes:
\be
ds^2 \simeq 
-\alpha ^2 z^2 \left(dX^0\right)^2 + dz^2 + \sum_{i=1}^{3}\left(dX^i\right)^2 
\label{conical}
\ee
implying that $X^0$ plays now the r\^ole of a {\it Minkowskian}-signature
temporal variable, despite its original Euclidean nature. 
This is a result of the identification $\phi=X^0$, and the fact that 
$\phi$ appeared with Minkowskian signature due to the supercriticality 
(cf. (\ref{centralcharge})) of the Liouville string under consideration. 

Notice that the space time (\ref{conical}) is flat, and hence it satisfies
Einstein's equations, formally. However,
the space time (\ref{conical}) has a {\it conical} singularity  
when one compactifies the time variable $X^0$ on a circle of finite radius
corresponding to 
an inverse `temperature' $\beta$. Formally, 
this requires a Wick rotation 
$X^0 \rightarrow iX^0$ and then compactification, $iX^0=\beta e^{i\theta}$,
$\theta \in \left(0 , 2\pi \right]$.
The space-time then becomes a {\it conical} space-time of Rindler type
\be
  ds_{conical}^2 = \frac{1}{4\pi^2}\alpha^2 \beta^2 z^2 \left( d\theta \right)^2 + dz^2 
+ \sum_{i=1}^{3}\left(dX^i\right)^2 
\label{conical2}
\ee 
with deficit angle $\delta \equiv 2\pi - \alpha \beta$. 
We recall that there is a 
`thermalization theorem' for this space-time~\cite{unruh}, in the 
sense that the deficit 
disappears and the space-time becomes regular, when the temperature 
is fixed to be 
\be 
     T = \alpha /2\pi  
\label{unruh}
\ee
The result (\ref{unruh}) may be understood physically 
by the fact that $\alpha$ is essentially related to recoil. As discussed
in \cite{kanti98}, the problem 
of considering a suddenly fluctuating (or recoiling) brane at $X^0=0$,
as in our case above, 
becomes equivalent to that of an observer in a (non-uniformly)
accelerated frame. At times long after the collision the acceleration 
becomes uniform and equals $\alpha$. This implies the appearance of a
non-trivial  
vacuum~\cite{unruh}, characterized by thermal properties of the form 
(\ref{unruh}). At such a temperature the vacuum becomes 
just the Minkowski vacuum, whilst the Unruh vacuum~\cite{unruh} 
corresponds to $\beta \rightarrow \infty$. 
Here we have derived this result in a
different way than in \cite{kanti98}, but the essential physics is 
the same. Notice that the presence of such a `thermal'
effect is a manifestation of the breaking of Lorentz
invariance induced by $D$-brane recoil. We study further
observational consequences of this effect in Section 7.

\begin{figure}[b]
\begin{center}
\includegraphics[width=.3\textwidth]{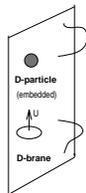}
\end{center}
\caption[]{The world as a $D3$ brane 
`punctured' by $D$ particles ($D0$ branes). 
The scattering on the $D0$ brane of string states, either
closed (gravitons) or open (matter fields) that
live on the $D3$ brane, cause the $D0$ brane
to recoil, leading to stochastic effects in the
propagation of the low-energy states, as well as to 
non-zero `vacuum' energy on the $D3$ brane.}
\label{fig3}
\end{figure}

\section{Vacuum Energy and Expansion of the Universe Induced by $D$-Brane
Recoil}

In the picture envisaged above, where our world is viewed as a 
fluctuating $D$--brane, one may consider more complicated configurations
of intersecting branes. The simplest of all cases is the one
depicted in Fig.~\ref{fig3}, in which a $D$ particle is embedded in
a Euclidean $D4$ brane, which is itself embedded in a higher-dimensional
(bulk) space-time.


In this case, any low-energy string state living on the $D3$ brane
which scatters off the emebedded $D0$ brane will cause
recoil of the latter and hence distortion of space-time, according to 
the above discussion. The distortion is such as to induce
a cosmological constant on the $D3$ brane, as discussed in 
detail in~\cite{emncosmol}. 
The simplest case to consider is that in which the embedded
defect is very heavy, so that one may ignore its recoil velocity
and consider only its quantum fluctuations, which are associated with the
uncertainty 
in its position, i.e., with the $C$ operator in (\ref{logpair}).
In that case, as discussed in~\cite{emncosmol}, 
the space-time surrounding the recoiling (fluctuating) $D$ particle 
becomes at first sight anti-de-Sitter~\footnote{To be precise,
the corresponding stringy $\sigma$ model becomes conformally equivalent
(i.e., up to marginal deformations on the world-sheet)
to a $\sigma$ model in an anti-de-Sitter target space-time.}, 
with a small negative cosmological constant of the form: 
\be
  V^{bulk}_{cosm~const}= -4 D(D-1) \epsilon^4 \cdot M_s ^4 
\label{cosmol}
\ee
where $D$ is the dimensionality of
the target space-time in which the $D$ particle is embedded, 
and $\epsilon^2 \sim {\rm Ln}\left(L/a\right)$ is related to the 
world-sheet size of the non-critical string, and thus to the 
renormalization-group scale, on account of (\ref{defeps}).  
 
However, upon the identification of this scale $\epsilon^{-2}$ with the 
(Euclideanized) 
target time $X^0$, one observes~\cite{emncosmol} 
that when one continues analytically the 
Euclidean time $X^0$ to a Minkowskian time variable $it$, the quantity
$\epsilon^2
\rightarrow i\epsilon^2 $. This implies that the `vacuum energy'
due to the recoiling $D$ particle in Fig.~\ref{fig3}
becomes {\it positive}, and relaxes to zero asymptotically as $1/t^2$.    
As discussed in~\cite{emncosmol} such a time dependence may
not be inconsistent with 
recent astrophysical observations~\cite{sn}. 

The identification of the scale $\epsilon^{-2}$ with the Euclideanized 
target time $X^0$ raises the question whether some variants of 
Einstein's equations are satisfied in this case. 
We note that, upon this identification, the quantity $\alpha$ in 
(\ref{conical}) depends on the time $t$, since by definition it
depends on $\epsilon$. Thus, the resulting metric is no longer 
flat. 
Even in this case, however, the satisfaction of Einstein's equations 
is guaranteed in the bulk space-time, as 
shown in~\cite{am}, provided there is a non-trivial dilaton field 
of the form: 
\begin{equation} 
\varphi \,\,\,\, \propto \,\,\,\, {\rm ln} \,\, t 
\label{dilaton} 
\end{equation}
In fact, as emphasized in~\cite{am}, an important r\^ole is played by 
averaging over quantum fluctuations in the position 
of the $D3$ brane along the bulk direction in the case of 
Fig.~\ref{fig3}. Such fluctuations are properly taken into 
account by summing up over world-sheet topologies, as
explained in~\cite{mavro+szabo}. 
This leads effectively to a dimensional reduction in the associated geometry,
given that in this picture the extra (bulk) dimensions
are viewed as couplings of a $\sigma$ model over which 
a four-dimensional observer averages~\cite{am}. 

If one denotes by $\sigma^2$ 
the corresponding uncertainty in the position of the $D3$ brane 
along the fifth (bulk) dimension in the geometry of Fig.~\ref{fig3},
then the analysis of \cite{am} has demonstrated that the 
distortion of the (four-dimensional) space-time 
due to such fluctuations, within the framework of the identification
of the scale $\epsilon^{-2}$ with the target time $t$, 
can be described by the following invariant line element:
\begin{equation}
<<ds^2>>^{(4)}=\frac{b^2\sigma^2}{t^2}(dt)^2 -
\left|1 - \frac{b^2\sigma^2}{t^2}\right|\sum_{i,j=1}^{3}\delta_{ij}dx^i dx^j,
~~b={\overline \Gamma}/2\sqrt{2}
\label{fourdimmetric}
\end{equation}
where ${\overline \Gamma}$ is related 
to the uncertainty in the momentum of the $D3$ brane, as 
defined in (\ref{gauss}). 

Notice that the product $b^2\sigma^2$ is just the 
momentum-position uncertainty product 
of the quantum-fluctuating $D3$ brane,
which is saturated by the uncertainty principle in its stringy version,
as discussed in~\cite{mavro+szabo}. Its precise value depends on the
form of the recoiling $D$-brane state, something which at present
cannot be known exactly. It is the lack of such knowledge that 
prevents us, at present, from determining dynamically the 
scale $\alpha$, and thus deriving dynamically (or excluding
phenomenologically!)
the induced hierarchy of mass scales. In our opinion,
this problem is still unsolved, 
given that even in the superstring/supermembrane scenaria 
of~\cite{randal99}, the induced low-energy scale is introduced in an 
{\it ad hoc} manner, constrained only by consistency
with phenomenology.

We now remark that, if one ignores the recoil fluctuations of the $D$ 
particle
in Fig.~\ref{fig3}, and considers only the 
fluctuations of the $D3$ brane, then the 
induced `vacuum energy' on the brane $\Lambda$ has the form:
\begin{equation}
\Lambda = \frac{(5b^2 - 8)}{t^2}M_s^4, 
\label{cosmolconst}
\end{equation} 
In addition to this field-independent vacuum energy, one also obtains a 
positive-definite excitation energy $V$ 
for the $D3$ brane, due to the fact that 
the recoiling $D$ brane finds itself  in an excited state rather 
than in its ground state. The result is~\cite{am}:
\begin{equation}
V = \frac{(2b^2 + 4)}{t^2}M_s^4, 
\label{excite}
\end{equation} 
where we
have passed to a Minkowskian target time $t$
in (\ref{cosmolconst}) and (\ref{excite}).

The analysis of \cite{am} showed that,
in the classical limit where 
the position uncertainty of the $D3$ brane vanishes, 
$\sigma \rightarrow 0$, the 
dilaton equation of motion forces a dynamical constraint on the 
width parameter $b^2=8/5$, which, {\it remarkably}, constrains  
the cosmological constant $\Lambda$ 
to vanish. Notably the above value of $b$ is compatible 
with the supercriticality of the $\sigma$ model,
which is essential for the self-consistency of the 
identification of the Liouville 
mode with the target time $X^0$. 
In the case where $\sigma \ne 0$, one may still demand the cosmological 
constant to vanish, but then this constrains the uncertainty 
$\sigma$ to a particular value  that nearly saturates
position-momentum uncertainty for the $D3$ brane: $\sigma=5/(\sqrt{96}M_s)
\simeq 1/(\sqrt{2M_s})$. We remark,
comparing the contributions (\ref{cosmolconst}, \ref{excite}) to the
vacuum and excitation energies
(\ref{cosmol}), 
that fluctuating defects on the $D3$ brane
in the geometry of Fig.~\ref{fig3} 
yield 
contributions to the vacuum energy on the brane that scale 
similarly with the target time, i.e., as $1/t^2$. 

In the case that the recoiling (momentum) fluctuations 
of the embedded $D$ particle are not ignored, one obtains 
additional contributions to the four-dimensional `vacuum energy'. 
To see this, we recall that the four-dimensional space-time,
on which the defect is embedded, is to be viewed as a bulk space-time
from the point of view of the world-sheet approach to the recoil of 
the $D$ particle. Following the same approach as that leading to 
(\ref{conical}), involving the identification of the 
Liouville field with the target time, $t$, 
one observes again that there exists an (expanding) horizon, located at 
$r^2 \equiv x^2 + y^2 + z^2 =t^2 /b'^2 $
where $b'$ is related to the momentum uncertainty of the fluctuating 
$D$ particle, and 
$\{ x_i \}, i=1,\dots 3$
constitute the bulk dimentions, obeying Dirichlet boundary conditions
from a world-sheet view point. 
For the region of space time
{\it inside the horizon} one 
obtains 
the following  metric 
on the $D3$ brane, as a result of recoil of the $D$ particle
embedded in it:
\be
ds^{2{(4)}} \simeq 
\frac{b'^2 r^2}{t^2} \left(dt\right)^2 -  
\sum_{i=1}^{3}\left(dx^i\right)^2~,\qquad r^2 = \sum_{i=1}^3 x_i^2  
< t^2/b'^2
\label{fourdime2}
\ee
We can show easily that this metric is a solution of 
Einstein's  equations in a four-dimensional space-time $\{x_i, t \}$,  
with a vacuum energy $-\Lambda$ that is constant in time, but 
position-dependent:
\be
\Lambda = \frac{2}{r^2}~; 
\label{coconst}
\ee
and a four-dimensional dilaton field of the form:
\be 
  \varphi = {\rm ln} \, r + b'{\rm ln} \, t  
\label{lineardila}
\ee
Incidentally, we remark that the scalar curvature corresponding to the metric 
(\ref{fourdime2}) 
has the form $R=-4/r^2$, and as such has a singularity 
at the initial 
location 
$r=0$ of the $D$-particle defect, as expected. 

We next notice,
from the forms of the metrics (\ref{fourdimmetric}) 
and (\ref{fourdime2}), that
the choice of time $t \sim \epsilon^{-2}$, i.e., directly 
proportional to the Liouville world-sheet zero mode,  is not 
the appropriate one for a Friedmann-Robertson-Walker (FRW) Universe, 
which describes to a good approximation the world around us. 
It is straightforward to see that the metric 
(\ref{fourdimmetric}) becomes of FRW type upon the transformation 
\begin{equation}
t \rightarrow t_{FRW} =b \, \sigma \, {\rm ln} \, t
\label{friedmanntime}
\end{equation} 
We then observe from (\ref{dilaton}) that 
the dilaton is {\it linear}~\cite{aben89} in this time coordinate: 
$\varphi \propto t_{FRW}/b\sigma $ and that
the scale factor increases as $|1 - b^2\sigma^2~e^{-2t_{FRW}/b\sigma}|$,
whilst the 
various contributions 
(\ref{cosmolconst}, \ref{excite}) to the vacuum and/or excitation energies 
become {\it exponentially
suppressed}. 

However, we also remark that there is a model in which 
the $1/t^2$ scaling pertains directly to the Friedmann-Robertson-Walker
time coordinate. This is the original case described in~\cite{emncosmol},
where the world was not viewed as a brane, but 
there were simply $D$ particles embedded in the four-dimensional
space-time. In such a case, the recoil 
of the $D$-particle defect is also described by a Liouville 
theory, but in that case the coefficient of the Liouville coordinate
is unity, which leads to a direct identification of the Liouville 
field with Friedmann-Robertson-Walker time.

\section{Energy Conditions and Horizons in Recoil-Induced Space-Times}

It is interesting to look at the energy conditions 
of such space times, which would determine whether
ordinary matter can exist in such regions. 
As is well known there are various forms of energy 
conditions~\cite{energycond}, which may be expressed as follows: 
\begin{eqnarray} 
{\rm Strong}~~~ &:&~~~ \left(T_{\mu\nu}-\frac{1}{D-2}
g_{\mu\nu}T_\alpha^\alpha\right)\xi^\mu\xi^\nu \ge 0, \nonumber \\
{\rm Dominant}~~~&:& ~~~ T_{\mu\nu}\xi^\mu\eta^\nu \ge 0, \nonumber \\
{\rm Weak}~~~&:& ~~~T_{\mu\nu}\xi^\mu\xi^\nu \ge 0, \nonumber \\
{\rm Weaker}~~~&:& ~~~T_{\mu\nu}\zeta^\mu\zeta^\nu \ge 0.
\label{energyconds}
\end{eqnarray}
where $g_{\mu\nu}$ is the metric, and $T_{\mu\nu}$ is the 
stress-energy tensor in a $D$-dimensional space time,
including vacuum energy contributions,   
$\xi^\mu,\eta^\mu$, are arbitrary future-directed time-like
or null vectors, and $\zeta^\mu$ is an arbitratry null vector. 
The conditions have been listed in decreasing strength, 
in the sense that
each condition is implied by all its preceding
ones. 

It can be easily seen from Einstein's equations for the metric
(\ref{fourdime2}) 
that inside the horizon $b'^2r^2 \le t^2$ the 
conditions are satisfied, which implies that stable matter can 
exist  {\it inside} such regions of the recoil space time. 
On the other hand, {\it outside the horizon} 
the recoil-induced metric assumes the form: 
\begin{equation}
ds^{2{(4)}} \simeq 
\left(2 - \frac{b'^2 r^2}{t^2}\right) \left(dt\right)^2 -  
\sum_{i=1}^{3}\left(dx^i\right)^2~,\qquad r^2 > t^2/b'^2
\label{fourdime2b}
\end{equation} 
The induced scalar curvature is easily found to be:
$$R=-4b'^2\left(-3t^2 + b'^2r^2 \right)/\left(-2t^2 + b'r^2\right)^2~.$$ 
Notice that there is a {\it curvature} singularity 
at $2t^2 =b'^2r^2$, which is precisely  
the point where there is a signature change in the metric 
(\ref{fourdime2b}).

Notice also that, 
in order to ensure 
a Minkowskian signature in the space-time (\ref{fourdime2b}),
one should impose the restriction  
\begin{equation} 
2 > \frac{b'^2r^2}{t^2} > 
1~; 
\label{signature}
\end{equation} 
It can be easily shown that the weaker energy condition (\ref{energyconds})  
can be satisfied 
for times $t$ such that 
\begin{equation} 
\frac{b'^2r^2}{t^2} \simeq 1+\varepsilon; \qquad \varepsilon \rightarrow 0^+
\label{encon}
\end{equation} 
i.e., on the initial horizon. 
To see this, it suffices to notice that the weaker energy condition
reads in this case:
\be 
     \left(4t^2 - b'^2r^2\right)\left(2 - \frac{b'^2r^2}{t^2}\right)(\zeta^0)^2
\le b'^2\left(\sum_{i=1}^{3} x^i \zeta^i \right)^2 
\label{weak2}
\ee 
where we used the fact that $\zeta^\mu$ is a null vector. 
Choosing $\zeta_1 \ne 0, \zeta_i=0, i=2,3$, it can be shown that
the right-hand-side of the above inequality can be bounded from above
by
\begin{equation}
b'^2 r^2 \sum_{i=1}^{3} (\zeta^i)^2=b'^2 r^2 \left(2-
\frac{b'^2r^2}{t^2}\right)
(\zeta^0)^2,
\end{equation}
which, on account of
the requirement (\ref{weak2}) would imply $\left(2t^2 -b'^2r^2\right)\le 0$.
This is in contradiction with the range of validity of (\ref{fourdime2b}),
unless one lies on 
the initial 
horizon (\ref{encon}). 
Notice that in this region of space-time there is a smooth matching 
between the interior (\ref{fourdime2}) and the exterior (\ref{fourdime2b}) 
geometries. In such regions 
of space time, surrounding the recoiling defect, 
matter can exist in a {\it stable form}.

The above considerations suggest that matter can  be trapped  
{\it inside} such horizon regions around a fluctuating $D$-particle defect. 
This sort of trapping is interesting
for our space-time-foam picture, as it implies that such
{\it microscopic $D$-brane  
horizons act in a similar way as black-hole horizons.}
The {\it dynamical} formation of horizons
that trap matter around a recoiling heavy $D$-particle defect is 
reminiscent of black-hole horizons obeying the cosmic censorship
hypothesis. 
Of course, this situation pertains 
to a single defect embedded in the four-dimensional
space-time. The
extension to a multi-defect situation, 
in which $D$-particle defects appear as quantum fluctuations 
rather than real defects, is still pending, even in a dilute-gas
approxiamtion. This would be 
closer to a realistic space-time foam picture~\footnote{The above formulae
assume that 
the defects are point-like.
However, if one views the defects as having 
small dimensions, which could come from the compactification of extra
dimensions as in the case of string theory, then 
the analysis of \cite{leonta} can be appropriately extended 
to this case 
in a way that does not qualitatively change the above conclusions.}.

\section{Supersymmetry Breaking Induced by $D$-Brane Recoil}

The thermalization of the region 
of the bulk space time $\alpha^2 z^2 << 1$
implied by (\ref{unruh}) also implies that any 
low-energy effective field theory of excitations that 
are constrained to live 
on the brane world $\{ X^I \}$ at $z=0$ will be 
exposed to this (small but) finite temperature.  
This has important implications for symmetry obstruction,
as discussed in \cite{adrian+mavro99,witten95}.
The recoiling $D$--brane world constitutes an excited state,
and the finite induced temperature breaks both the Lorentz symmetry and
the supersymmetry (if the latter is assumed) of the low-energy effective
theory. The symmetry breaking is 
an obstruction, in the sense of \cite{witten95}, rather than 
a spontaneous breaking, because the $D$ brane is not in its ground state.
Notice that the mathematical origin of the obstruction may be traced 
back to the impossibility of defining globally covariantly-constant
spinors in space-times with deficits~\cite{henneaux}. 
The induced supersymmetry breaking may be studied  
explicitly using the `thermal superspace' (TS) formalism 
of~\cite{derend}. For simplicity, we review here only
the basic results of this formalism that are relevant for our purposes.
The interested reader is advised to study~\cite{derend}
for further details. 
 
A point in TS is described by, in addition to the usual 
space-time coordinates, Grassmann parameters 
that, in contrast to the 
zero-temperature case, are {\it time-dependent} and 
{\it antiperiodic} in the imaginary time on the interval 
$\left(0, \beta \right]$:
\be
  {\widehat X} = \left( {\widehat x}^\mu, {\widehat \theta}^\alpha (t), 
{\widehat {\overline \theta}}^{\dot \alpha} (t) \right)
\label{tspace}
\ee
Here the $\widehat{X}$ notation denotes TS quantities, and 
\be
    {\widehat \theta}^\alpha (t + i \beta) = -{\widehat \theta}^\alpha (t)~,
\qquad {\widehat {\overline \theta}}^{\dot \alpha} ( t + i \beta) =
-{\widehat {\overline \theta}}^{\dot \alpha} ( t )~.
\label{antiperiod}
\ee
Due to this time dependence in the parameters ${\widehat \theta}$, 
the spinor parameters of the infinitesimal 
transformations ${\widehat \epsilon} $ in thermal supersymmetry 
are also {\it time-dependent}, in contrast to the zero-temperature 
supersymmetric case, where such spinors are space-time constants. 
For instance, 
a scalar superfield ${\widehat \phi} (x, {\widehat \theta}, 
{\widehat {\overline 
\theta}})$ transforms in a way that is formally analogous 
to the $T=0$ case~\cite{derend}, but with the important 
difference that $\epsilon$ is now time-dependent and antiperiodic 
in temperature:
\be
   \delta {\widehat \phi} =i \left({\widehat \epsilon}^\alpha 
{\widehat Q}_\alpha + {\widehat {\overline \epsilon}}_{\dot \alpha}
{\widehat {\overline Q}}^{\dot \alpha}\right)~, \qquad 
{\widehat \epsilon}(t + i\beta)=-{\widehat \epsilon}(t)
\label{susytrns}
\ee
It should be stressed that 
the time dependence of the supersymmetry parameters 
is essential for the breaking of the 
thermal supersymmetry at finite temperature~\cite{derend}, 
and, essentially, arises 
from the difference in boundary conditions 
between bosons and fermions. 

Using such a TS formalism, one may calculate {\it thermal mass splittings}
for the various supermultiplets. For instance, in the case of 
the four-dimensional Wess-Zumino model, examined in \cite{derend},
the mass splittings at $T \ne 0$
between the bosonic and fermionic degrees of freedom in the model
are of the form: 
\be
   M_{3,n}^B=M_4^2 + \frac{4\pi^2 n^2}{\beta^2}. \qquad 
   M_{3,n}^F=M_4^2 + \frac{\pi^2 (2n + 1)^2}{\beta^2}. \qquad 
\label{split}
\ee
where $M_4$ is the $T=0$ mass of the Wess-Zumino model.
The notation $M_{3,n}^{B,F}$ denotes the thermal mass of the $n$'th mode
for bosons (B) or fermions (F) 
in the effective three-dimensional Euclidean 
theory at temperature $T=\beta^{-1}$, obtained after the 
thermal expansion. 

It is clear from (\ref{split}) that the mass degeneracy of the $T=0$
(rigid) supersymmetry is lifted, signalling 
{\it thermal supersymmetry breaking} at the level of the thermal fields. 
Such a breaking is also manifest in the finite-temperature action
of the 
model, in the sense that the thermal supersymmetric variations of 
the action are proportional to the time-derivative 
$\partial _t {\widehat \epsilon}$ (in a Wick-rotated time),
which is non-trivial. Expressed in terms of Matsubara thermal modes,
the total variation of the action is then found to be proportional to 
$\omega_n^F\equiv \pi^2 \beta^{-2} (2n + 1)^2 \sim T$, 
as a consequence of thermal supersymmetry breaking.
Obviously, in the limit $T \rightarrow 0$, the thermal breaking of 
supersymmetry evaporates.

We now come to our $D$--brane case, where, as we have seen, 
there is a `thermalization' of the 
space-time surrounding a recoiling $D$ brane (\ref{unruh}).
The lifting of the mass degeneracy discussed above 
results in a non-trivial contribution to the 
vacuum energy. 
It should be noted that the $D$--brane recoil deformations (\ref{logpair}) 
are essentially higher-genus effects in string theory, as 
discussed in detail in \cite{emn98}. Such effects are 
target-space   
quantum effects. In ordinary field theories, it is well known how
quantum effects contribute to the effective potential,
which has crucial consequences for the so-called gauge hierarchy
problem in field theories with spontaneous breaking of a gauge symmetry.
For (rigid) supersymmetric theories, for instance, one-loop 
contributions to the 
effective potential have the generic form~\cite{iliopoulos,kounnas}: 
\be
V = V_0 + \frac{1}{64\pi^2}{\rm Str}{\cal M}^0 \cdot \Lambda ^4 \cdot 
{\rm ln}\frac{\Lambda^2}{\mu^2} + \frac{1}{32\pi^2}
{\rm Str}{\cal M}^2 \cdot \Lambda^2 + 
\frac{1}{64\pi^2}{\rm Str}{\cal M}^4 \cdot {\rm ln}\frac{{\cal M}^2}{\Lambda^2}
+ \dots 
\label{effpot}
\ee
where $\Lambda$ is an ultraviolet 
cut-off for the low-energy effective field theory, 
which, in our context,  may be taken to be of order of the string scale
$M_s$, $V_0$ is the classical potential, the 
$\dots $ denote $\Lambda$-independent 
contributions, $\mu$ is a scale parameter, and the Supertrace
${\rm Str}{\cal M}^n$ is defined as:
\be
   {\rm Str}{\cal M}^n \equiv \sum_{i} (-1)^{2J_i} \left( 2J_i + 1
\right)m_i^n 
\label{supertrace}
\ee
where the $m_i$ are the mass eigenvalues of the various field species,
and the $J_i$ are their spins. 
In rigid supersymmetric theories, ${\rm Str }{\cal M}^n =0$ 
as a result of the mass
degeneracy within the supermultiplets. 
This is not the case for $n > 0$, if supersymmetry is broken
dynamically. 
In our case of $D$--brane recoil (\ref{split}),
there is a recoil-induced dynamical
split between the modes, due to the breaking of the 
thermal supersymmetry. In this case the various supertraces
involve sums over thermal modes which need regularization. 
Typically, due to the fermionic contributions, there are $\zeta (0)$ 
terms appearing, where $\zeta$ is the Riemann zeta function.
For instance, from (\ref{split}) one would expect 
qualitatively a contribution to the effective potential 
in the Euclidean $d=3$-dimensional theory, after 
thermal expansion, of the form: 
\be 
     V \ni \frac{M_{uv}^2}{32\pi^2}{\rm Str}{\cal M}^2 \sim  
\sum_{i=1}^{N_s} \frac{b^2M_{uv}^2}{8\pi^2t^2}\zeta (-1,\frac{1}{4})=  
-\frac{N_s M_{uv}^2b^2}{384\pi^2 t^2} 
\label{zetas}
\ee
where $N_s$ is the number 
of particle species, and $M_{uv}$ is an ultraviolet cut-off mass scale 
in the low-energy effective theory. It it natural to identify 
it with the string (or quantum gravity) scale $M_s$~\footnote{In our
approach we take $M_s$ to be high, near the Planck scale, and hence we do 
not distinguish between the gravity and string scales.}. 
We stress that 
such contributions should be attributed 
to loop corrections rather than to the classical potential, 
as they are associated with the higher-genus nature of 
the recoiling operators (\ref{logpair}), as explained in 
\cite{emn98} and mentioned above. 

It is an important issue whether, if one selects $T \sim \alpha \sim
{\cal O}(1) {\rm TeV}
<< M_{Planck} \sim 10^{19}$ GeV, the supersymmetric `solution' to the
hierarchy problem   
is stable to higher orders in the loop expansion. 
For theories with softly-broken rigid supersymmetry
this is guaranteed, as the corrections to the Higgs mass coming from 
the logarithmic terms in (\ref{effpot}) are at most of order 
${\cal O}(M_{SUSY}^2)$~\footnote{We recall that in supersymmetric theories
the $\Lambda ^4$ terms 
are absent, since there are equal numbers of fermions and bosons in each
multiplet, implying that ${\rm Str}{\cal M}^0=0$.}. 
However, the situation changes drastically in supergravity theories.
In general, for local supersymmetry spontaneously broken 
at a scale $m_{SUSY}$,
a gravitino mass $m_{3/2} \sim m_{SUSY} \sim 1$ TeV $ 
<<  M_{Planck}$ does not necessarily guarantee that the 
hierarchy is stable under quantum corrections~\cite{kounnas}.  

In our string $D$--brane--inspired model we naturally 
have supergravity in the brane world $\{ X^I \}$,$I=0, \dots 3$.
In such a case a detailed thermal expansion,
using and extending TS techniques to include 
supergravity models~\cite{derend}, should be performed before 
making any 
quantitative statement about the phenomenological consequences 
of the recoil-induced supersymmetry breaking on the brane world. 
However, some qualitative statements may be made.
For instance
it is natural to ask (at least 
qualitatively) whether the hierarchy implied by the 
presence of the scale $\alpha << M_s$, with the string scale $M_s
\sim 10^{18}$ GeV, 
is stable against quantum corrections.  

The interesting issue is whether realistic phenomenology can be achieved
(at least within the one-loop approximation made in the above treatment 
of recoiling $D$--brane world), in the sense that 
the resulting contributions to the vacuum energy
are compatible with astrophysical data 
and with the scale $\alpha \sim \epsilon \Gamma 
\sim {\cal O}(1)$ {TeV}
expected in the prospective resolution of the gauge hierarchy problem.
This depends crucially on our  interpretation of $\epsilon^{-2}$ 
as the target time. Indeed, one should note that 
the width $\Gamma$ of the Gaussian fluctuations of the recoil momenta 
$\Delta P^i$ entering $\alpha M_s $, see
(\ref{bendinglineel3ab}),  
satisfies an uncertainty relation~\cite{mavro+szabo},
$\Delta P_i \Delta Y_i \ge \hbar + {\cal O}(\ell_s^2 \Delta P_i^2) $,
and $\Delta Y_i \ge g_s^{1/3} \ell_s + \dots $,  
where $g_s$ is the string coupling and 
$\ell_s \sim M_s^{-1}$ is the string length. Thus 
$\Delta P_i $ can be arbitrarily bigger than $M_s$, 
if, for instance, the excited (recoiling) $D$--brane 
is localized in the bulk space-time 
at a distance of order $\ell_s$. In principle, of course, 
just exactly how big $\Delta P_i$ would be 
is determined by the appropriate wave functional
of the brane universe, which is beyond any 
detailed description at present. But phenomenological 
assumptions may actually shed light on such an important 
issues in quantum gravity.   

In principle, one can arrive in our scenario for space-time foam
at a situation where $\alpha \sim \epsilon^2 {\overline \Gamma}  
\sim 1$ TeV, as required by the phenomenological constraints on 
supersymemtry breaking and the solution to the hierarchy problem, 
whilst the vacuum energy is compatible with 
the current observations, simply because there may be cancellations 
among the various components (\ref{cosmolconst}, \ref{excite},
\ref{cosmol}) and (\ref{zetas}), which 
come with different signs, as we have seen. Such deep issues require
serious investigations before definite
conclusions are reached, but we believe this approach deserves further
study.

\section{Observable Breakdowns of Lorentz Invariance}

In this Section we ignore the possible quantum fluctuations of the 
$D3$ brane in the geometry of Fig.~\ref{fig3}, and concentrate only
on the quantum fluctuations of the $D$ particles embedded
in the $D3$ brane. This is equivalent to 
the original case discussed in our earlier work on the 
subject~\cite{ellis98,kanti98}, 
according to which the observed four-dimensional space-time is just
punctured 
with real $D$-particle defects, without being viewed as a $D$-brane. 

One of the most important consequences of the induced metric (\ref{gemetric})
is the induced refractive 
index for massless probes~\cite{sarkar} propagating in 
such space-times, which we now discuss in more detail. 
It is well known that light propagating through media with
non-trivial optical properties may
exhibit a frequency-dependent refractive
index, namely a variation in the light
velocity with photon energy. Another possibility is
a difference between the velocities of light with
different polarizations, namely birefringence,
and a third is a diffusive spread in the apparent velocity of
light for light of fixed energy (frequency). Within the
framework leading to the induced metric (\ref{gemetric}),  
the first~\cite{aemn} and third~\cite{emn98,ellis99} 
effects have been derived via a
formal
approach based on a Born-Infeld Lagrangian using
$D$-brane technology~\footnote{The possibility 
of birefringence has been raised~\cite{pullin} within a
canonical approach to quantum gravity, but we do not pursue such a
possibility here.}. A different approach
to light propagation has been taken in~\cite{ford}, where
quantum-gravitational fluctuations in the light-cone have been
calculated. Here we use this formalism 
to derive a non-trivial refractive index and a diffusive spread
in the arrival times of photons of given frequency. 

We first review briefly the analysis in \cite{ford},
which considered gravitons in a squeezed coherent state, the natural result 
of quantum effects in the presence of black holes.  
Such gravitons induce quantum fluctuations in the space-time metric, 
in particular fluctuations in the light-cone~\cite{ford},
i.e., stochastic fluctuations in the 
velocity of light propagating through this `medium',
Following~\cite{ford},
we consider a flat background space-time 
with a linearized perturbation, corresponding to the 
invariant metric element
$ds^2=g_{\mu\nu}dx^\mu dx^\nu = 
\left(\eta_{\mu\nu} + h_{\mu\nu}\right)dx^\mu dx^\nu 
= dt^2 - d{\overline x}^2 + h_{\mu\nu}dx^\mu dx^\nu $.
Let 
$2 \sigma (x,x')$ be the squared geodesic separation 
for any pair 
of space-time 
points $x$ and $x'$, and let $2 \sigma_0(x,x')$ denote the corresponding 
quantity in a flat space-time background. 
In the case of small 
gravitational perturbations about the flat background,
one may expand
$\sigma = \sigma _0 + \sigma_1 + \sigma_2 + \dots$,
where $\sigma_n$ denotes the $n$-th order term in 
an expansion in the gravitational perturbation $h_{\mu\nu}$.
Then, as shown in~\cite{ford}, 
the root-mean-square (RMS) deviation from the classical propagation 
time $\Delta t$ is related gauge-invariantly~\cite{ford} to $<\sigma^2>$ 
by
\be
   \Delta t = \frac{\sqrt{<\sigma^2> -
<\sigma_0^2>}}{L} \simeq \frac{\sqrt{<\sigma_1^2>}}{L} + \dots 
\label{deltat}
\ee
where $L = |x' - x|$ is the distance between the source and the detector.

As commented earlier, one may expect Lorentz invariance to be
broken in a generic theory of quantum gravity, and specifically
in the recoil context discussed earlier in this paper. In the
context of string theory, violations of Lorentz invariance entail the
exploration of
non-critical string backgrounds, since Lorentz invariance is 
related to the conformal symmetry that is a
property of critical strings. As we have discussed,
a general approach to the formulation of
non-critical string theory involves introducing a 
Liouville field~\cite{distler89} as a conformal factor
on the string world sheet, which has non-trivial dynamics and
compensates the non-conformal behaviour of the string background,
and we showed in the specific
case of $D$ branes that their recoil after
interaction with a closed-string state
produces a local distortion of the surrounding 
space-time (\ref{gemetric}). 

Viewed as a perturbation about a flat target space-time, 
the metric (\ref{gemetric}) implies that 
the only non-zero components of $h_{\mu\nu}$
are:
\be
h_{0i} = \epsilon ^2 {\overline u}_i t \Theta _\epsilon (t)
\label{pert2}
\ee
in the case of $D$-brane recoil.
We now
consider~\cite{ellis99} light propagation along the $x$ direction in
the presence of a $D$-brane-induced metric fluctuation $h_{0x}$
(\ref{pert2})
in flat space, along a null geodesic
given by $(dt)^2 = (dx)^2 + 2 h_{0x} dt dx $.
For large times $t \sim {\rm Log}\Lambda/a
\sim \epsilon^{-2}$~\cite{kanti98}, $h_{ox} \sim
{\overline u}$, and thus we obtain
\be
\frac{cdt}{dx}={\overline u} + \sqrt{1 + {\overline u}^2} \sim 
1 + {\overline u} + {\cal O}\left({\overline u}^2\right)
\label{refr}
\ee
where the recoil velocity $\overline u$ is in the direction of the
incoming
light ray. Taking into account energy-momentum 
conservation  in the recoil process, 
which has been derived in this formalism as mentioned previously, 
one has a typical order of magnitude 
${\overline u}/c ={\cal O}(E/M_Dc^2)$,
where $M_D =g_s^{-1}M_s$ is the $D$-brane mass scale, with $M_s \equiv 
\ell_s^{-1}$. Hence (\ref{refr}) implies 
a subluminal energy-dependent velocity of light~\cite{ellis99,sarkar}:
\be
c(E)/c=1 -{\cal O}\left(E/M_Dc^2\right)
\label{vellight}
\ee
which corresponds to a {\it classical} refractive index.
This appears because the metric perturbation (\ref{pert2}) 
is energy-dependent,
through its dependence on ${\overline u}$.

The subluminal velocity (\ref{vellight})
induces a delay in the arrival of a photon of energy $E$
propagating over a distance $L$
of order: 
\be
      (\Delta t)_r =  \frac{L}{c}{\cal O}\left(\frac{E}{M_Dc^2}\right)
\label{figmerclass}
\ee
This effect can be understood physically from the fact  
that the curvature of space-time induced by the recoil is
${\overline u}-$ and hence energy-dependent.
This affects the paths of photons
in such a way that more energetic photons see more
curvature, and thus are delayed with respect to low-energy ones.

The absence of superluminal light propagation
was found previously via the formalism of the Born-Infeld 
lagrangian dynamics of $D$ branes~\cite{mavro+szabo,emn98}.
Furthermore, the result (\ref{figmerclass}) is in agreement 
with the analysis of~\cite{aemn}, 
which was based on a more abstract analysis of Liouville strings.
It is encouraging that this result appears also in the more conventional 
general-relativity approach~\cite{ford}, in which the underlying
physics is quite transparent.

In addition to the above mean-field effects, 
there are {\it stochastic} fluctuations
about this mean value, which 
arise from the summation over world-sheet topologies. 
The calculation for the $D$-brane recoil case yields~\cite{emn98}:
$\Delta t = {\cal O}\left(g_s\frac{EL}{M_Dc^3}\right)$,  
where the suppression by the extra power of $g_s$, as compared with 
the mean-field effect (\ref{vellight}), 
is due to the fact that the phenomenon is associated with
higher-order string loops. We
note that this second effect is {\it not} associated with 
any modification of the dispersion relation of the particle probes,
but pertains strictly to fluctuations in the arrival 
times of photons~\cite{emn98,ford}. 
By construction, the effect
is associated with energy-dependent quantum fluctuations~\cite{ellis99}
around the mean 
value of the recoil velocity along the incident direction.

\section{Observational Limits from Data on Gamma-Ray Bursters}

\begin{figure}[b]
\begin{center}
\includegraphics[width=.3\textwidth]{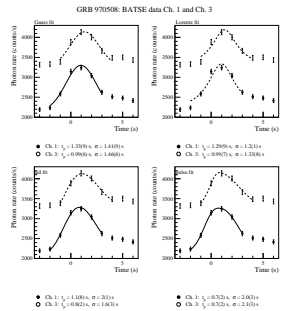}
\end{center}
\caption[]{Time distribution 
of the number of photons 
observed by BATSE in Channels 1 and 3 for GRB~970508, compared 
with the following fitting functions for the observed peaks~\cite{efmmn}:
(a) Gaussian, (b)
Lorentzian, (c) `tail' function, and (d) `pulse' function.
We list below each panel the positions $t_p$ and widths $\sigma_p$
(with statistical errors) found for each peak in each fit. We
recall that the BATSE data are binned in periods of 1.024~s.}
\label{figdat1}
\end{figure}

\begin{figure}[b]
\begin{center}
\includegraphics[width=.3\textwidth]{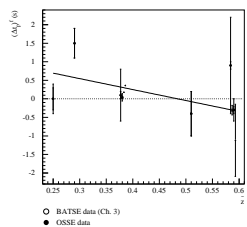}
\end{center}
\caption[]{Values of the shifts $(\Delta t_p)^f$ in
the timings of the emission peaks fitted for each GRB studied
using BATSE and OSSE data, plotted versus 
${\tilde z}=2 ( 1-(1+z)^{-1/2})$, where $z$ is the 
redshift. The indicated errors are the
statistical errors in the `pulse' fits provided by the fitting routine,
combined with systematic error estimates obtained
by comparing the results obtained using the `tail' fitting
function. The values obtained by comparing OSSE with BATSE 
Channel 3 data
have been rescaled by the factor $(E_{min}^{BATSE~Ch.~3} -
E_{max}^{BATSE~Ch.~1}) / (E_{min}^{OSSE} - E_{max}^{BATSE~Ch.~3})$,
so as to make them directly comparable with the comparisons of
BATSE Channels 1 and 3. The solid line is the best linear fit.}
\label{figregrz}
\end{figure}

Such speculations may be tested directly using experimental data
that are already available. Of particular interest are astrophysical
sources of energetic photons that are very distant and have short
time structures~\cite{sarkar}, such as Gamma-Ray Bursters (GRBs),
Active Galactic Nuclei (AGNs)~\cite{AGN} and pulsars~\cite{pulsar}. A
model analysis 
was presented in presented in~\cite{efmmn}, where we  
conducted a study of astrophysical data for 
a sample of GRBs
whose redshifts $z$ are known. Fig.~\ref{figdat1}
shows the data from a typical burster, GRB~970508.
We looked (without success) 
for a correlation with the redshift,
calculating the regression measure shown in Fig.~\ref{figregrz} for the 
effect (\ref{vellight}) and also its stochastic counterpart~\cite{efmmn}. 
Specifically, we looked for linear dependences of
a possible delay $\Delta t$ (and spread 
$\Delta \sigma$) in the arrival times of photons with higher energies,
proportional to $E/M$ ($E/\Lambda$) and ${\tilde z}
\equiv 2 \cdot [ 1 - (1 / (1 + z)^{1/2}] \simeq z - (3/4) z^2 +
\dots$, where the mass scales $M$ and $\Lambda$ might be associated with
quantum gravity. In the absence of any such effect,
we determined limits on $M$ and $\Lambda$ 
by constraining the possible magnitudes of the slopes  in 
linear-regression analyses of the differences between the arrival times
and widths of pulses in different energy ranges from five GRBs with
measured redshifts,
as functions of ${\tilde z}$.
Using the current value for the Hubble expansion parameter,
$H_0 = 100 \cdot h_0$~km/s/Mpc, where $0.6 < h_0 < 0.8$,
we obtained 
the following limits~\cite{efmmn}
\be 
M \gappeq 10 ^{15} \; {\rm GeV}, \quad \Lambda \gappeq 2 \times 10 ^{15} \;
{\rm GeV} 
\label{limit}
\ee
on the possible quantum-gravity effects.

We stress that,
as emphasized in \cite{efmmn}, it is only after a statistically
significant population of GRBs with measured redshifts are studied,
and possible systematic effects are excluded, that any
safe conclusions
about quantum gravity effects could be reached.
In this respect, searches for quantum-gravity effects using future
satellites such as GLAST and AMS~\cite{GLAST,AMS}  
would be essential.

Before closing this section, we would like to address 
some recent comments~\cite{mestres} on the refraction-index
effect (\ref{vellight}) we have discussed above. 
As pointed out in~\cite{mestres}, this effect may be probed
using cosmic rays, leading perhaps to the disappearance of the
GZK cutoff on Ultra-High-Energy (UHE) cosmic rays. Since this cutoff is
not seen in the data~\cite{olinto}, this may be observationally
acceptable. Similarly, it has recently been pointed out~\cite{PM} that
there
should have been an analogous suppression of 20~TeV photons from the
active galaxy Markarian 501, which seem, however, to escape being
cut off by the universal infrared background. The persistence of
20~TeV photons could perhaps be explained by Lorentz breaking of
the form $ \sim E / M_P$.

However, as also pointed out in~\cite{mestres}, there is an
issue with decays of lower-energy cosmic-ray particles.
We observe that there are two additional effects that need
to be taken into account in any analysis of such phenomena.
One is the possibility/likelihood of stochastic spreading in
the velocities of different particles  with the same energy,
discussed earlier in this paper. Another point is that the
possible effects of quantum gravity on decay vertices and interaction
processes in ultra-relativistic conditions have not been explored
theoretically. We find it quite plausible that more surprises may await us
here, and therefore reserve judgement on this issue, 
limiting ourselves for now to noting
that cosmic rays may also provide a valuable window on quantum-gravity
effects.

It has also been suggested~\cite{nu} that neutrino oscillation experiments
might also be sensitive to quantum gravity effects. The
interpretation of these experiments depends crucially whether
there are important flavour differences in any departure from
the conventional $E \sim p$ dispersion relation. If not, the
neutrino oscillation experiments would be insensitive to such a
quantum-gravity effect. On the other hand, any stochastic effect
on neutrino propagation could have an important impact on
neutrino-oscillation effects, and remains to be analyzed.

\section{Outlook} 

In this talk we have discussed a world-sheet approach to the 
problem of deformations of space times induced by 
the recoil of $D$-brane defects embedded in them
during scattering with low-energy string matter
(the latter may be real or virtual). 
As we have stressed here, as well as in our 
previous works on the subject, the recoil deformations
are not simply world-sheet boundary effects, as one may naively think, 
but they are associated non-trivially with bulk world-sheet
deformations of the corresponding stringy $\sigma$ models.
As a result, conformal invariance in the bulk of the world-sheet
is disturbed by their presence, and a standard Liouville 
theory dressing becomes necessary for its restoration.
It is this principle that determines in a unique way the form of the 
induced deformation in the target-space time surrounding the 
recoiling $D$-brane defect.

There are several issues that emerged from our analysis above.
The most important of them is the fact that the recoil induces
a Rindler structure in the space-time surrounding the 
$D$-brane defect.  
A study of the generic structure of string theories
in Rindler spaces is, from a formal view point, 
a highly-non trivial task. 
Some critical string theories in Rindler spaces, for given (discrete) values 
of the acceleration,  
can be represented 
as orbifold models~\cite{strom}. 
In our case, however, the value of the acceleration $\alpha$ 
varies continuously with the (world-sheet) renormalization scale,
and, as we have seen, a consistent treatment requires the 
(non-conformal) Liouville formalism. 
In general,
string theories in Rindler-like spaces, or space-times with deficits, 
are known to 
have instabilities (tachyons) in their spectrum,
even in the supersymmetric case~\cite{strom}, 
but this in our case is not a drawback, 
if one takes into account the excited state of the 
recoiling brane world and its relaxation towards equilibrium.

The presence of such `defective' space-times results in the 
obstruction of Lorentz symmetry, as well as 
of rigid (global) supersymmetry at a scale of order of the 
Rindler `acceleration' $\alpha$. 
The symmetry breakdown is an {\it obstruction} 
rather than a 
spontaneous breaking, as the recoiling $D$-brane
is in an excited state rather than its ground state.
At present we cannot determine exactly
the magnitude of $\alpha$: this is a dynamical feature 
of the underlying string theory, and hence can only be determined
by knowledge of the pertinent string/$D$-brane wavefunctional,
which is still lacking.

Another important feature of the recoil-induced space-times
is the presence of horizons {\it inside} which (stable) 
matter is trapped. 
Such horizons have great similarities with 
microscopic space-time boundaries in  
space-time foam situations.  

We have also shown that the recoil-induced space-times
possess non-trivial contributions to the vacuum energy, which  
relax to zero asymptotically in time.
In order to carry out  
realistic phenomenology, it would be necessary to determine
exactly the pertinent contributions to the `vacuum 
energy' on the brane from the   
various matter fields, which are much more complicated
in phenomenologically 
realistic models than the simplified situation studied above.
In general, the presence of matter excitations
on the (recoiling or quantum fluctuating) brane 
may either increase the supersymmetry-breaking scale 
or reduce the net 
contribution to the vacuum energy. 
These issues are left for future work, but we believe that 
the present considerations motivate further detailed
studies along this direction. 

Moreover, $D$-brane recoil appears to
induce non-trivial 
optical properties of space-time, such as a non-zero refractive index 
and stochastic propagation effects. Such effects 
are all  consequences of the obstructed Lorentz symmetry, 
and they modify the propagation 
of matter through such space-times with
consequences that may be observable in the foreseeable future, if
the light velocity is modified to ${\cal O}(E / M)$ where
$M \sim M_P$.
We have demonstrated that this is possible 
in the examples studied in 
this work.
Some of the observable consequences, namely those which are   
based on observations of light from GRBs and AGNs,
have been discussed briefly in this work.
Such analyses set at present the limit
for the minimum scale of these 
gravitational effects to $M \gappeq {\cal O}(10^{15})$ GeV. 

It should be stressed that, at present, the field is 
still in its
infancy, and may even turn out to be still-born. The mathematical models
we have discussed,
although consistent, are far from being realistic. 
However, we believe that the results obtained so far are 
sufficiently encouraging to motivate 
further studies in the future. There is even the possibility
of experimental verification (or exclusion) of such models,
so `quantum gravity phenomenology' may not be an oxymoron.

\section*{Acknowledgements}

We thank G. Amelino-Camelia, A. Campbell-Smith, K. Farakos, 
V. Mitsou and S. Sarkar
for enjoyable collaborations on work reviewed here.
The work of N.E.M. is partially supported by PPARC (UK)
through an Advanced Fellowship. 
That of D.V.N. is partially supported by DOE grant 
DE-F-G03-95-ER-40917.
N.E.M. and D.V.N. also thank
H. Hofer for his interest and support.

%

\end{document}